# Incommensurate antiferromagnetic order in weakly frustrated two-dimensional van der Waals insulator $CrPSe_3$

Baithi Mallesh[1,2], Ngoc Toan Dang[3,4,*], Tuan Anh Tran[5], Dinh Hoa Luong[1,2], Krishna P. Dhakal[2], Duhee Yoon[1], Anton V. Rutkauskas[6], Sergei E. Kichanov[6], Ivan Y. Zel[6], Jeongyoung Kim[2], Denis P. Kozlenko[6,*], Young Hee Lee[1,2,7,*], Dinh Loc Duong[1,2,*]

[1] Center for Integrated Nanostructure Physics, Institute for Basic Science, Suwon 16419, Republic of Korea

[2]Department of Energy Science, Sungkyunkwan University, Suwon 16419, Republic of Korea

[3]Institute of Research and Development, Duy Tan University, 550000 Danang, Vietnam

[4]Faculty of Environmental and Natural Sciences, Duy Tan University, 550000 Danang, Vietnam

[5]Ho Chi Minh City University of Technology and Education, 700000 Ho Chi Minh, Vietnam

[6]Frank Laboratory of Neutron Physics, Joint Institute for Nuclear Research, 141980 Dubna, Russia

[7]Department of Physics, Sungkyunkwan University, Suwon 16419, Republic of Korea

[*]Correspondence to: dangngoctoan1@duytan.edu.vn, denk@nf.jinr.ru, leeyoung@skku.edu, l.duong1902@gmail.com

**Abstract**

Although the magnetic order is suppressed by a strong magnetic frustration, it is maintained but appears in complex order forms such as a cycloid or spin density wave in weakly frustrated systems. Herein, we report a weakly magnetic-frustrated two-dimensional van der Waals material $CrPSe_3$. Polycrystalline $CrPSe_3$ was synthesized at an optimized temperature of 700 $^oC$ to avoid the formation of any secondary phases (e.g., $Cr_2Se_3$). The antiferromagnetic transition appeared at $T_N$ ~ 126 K with a large Curie–Weiss temperature $\theta_{CW}$ ~ –371 K via magnetic susceptibility measurements, indicating weak frustration in $CrPSe_3$ with a frustration factor $f$ ($|\theta_{CW}|/T_N$) ~ 3. Evidently, the formation of long-range incommensurate spin-density wave antiferromagnetic order with the propagation vector **k** = (0, 0.40, 0) was revealed by neutron diffraction measurements at low temperatures (below 120 K). The monoclinic crystal structure of C2/m symmetry is preserved over the studied temperature range down to 20 K, as confirmed by Raman spectroscopy measurements. Our findings on the spin density wave antiferromagnetic order in two-dimensional (2D) magnetic materials, not previously observed in the $MPX_3$ family, are expected to enrich the physics of magnetism at the 2D limit, thereby opening opportunities for their practical applications in spintronics and quantum devices.

## I. INTRODUCTION

Two-dimensional (2D) van der Waals (vdW) magnetic materials are increasingly being recognized for their potential application in next-generation spintronic devices [1–5]. These materials in atomically thin forms give rise to controlling magnetic order and physical properties by gate bias, strain, and proximity effects [3–10]. To explore novel materials and the physics of magnetism at the 2D limit, both intrinsic-magnetic and magnetic-doped 2D vdW materials have been synthesized and investigated [5,11–19]. Whereas the latter class requires sophisticated efforts in designing and investigation, the exploration of intrinsic 2D vdW magnetic materials relies on the available bulk forms like ferromagnetic insulating $CrBr_3$ [20] and $CrI_3$ [16], semiconducting $CrGeTe_3$ [15], metallic $Fe_3GeTe_2$ [21], and antiferromagnetic insulating $MPX_3$ (M = Fe, Mn, Ni, X = S, Se) systems [14,22–24]. In the most of these materials, the long-range magnetic order is retained in atomically thin forms, and in some cases like $CrI_3$, the inter layer magnetic interactions can be tuned from FM to AFM by mediating the layers number [16]. Furthermore, the antiferromagnetic order in $MPX_3$ can assist in the formation of coherent excitons [25,26], indicating an appropriate family of materials to investigate the Bose-Einstein condensation of excitons [26]. These materials are also suitable prototypes to search for high-$T_C$ superconductors by doping or using pressure [27].

Another important feature of the $MPX_3$ vdW materials is a diversity of commensurate antiferromagnetic orders, realized on structurally similar lattices with a hexagonal-like arrangement of transition metal atoms in the vdW layers (Table I). The magnetic structures can be Néel or zigzag with Heisenberg ($MnPS_3$ [28–31]), XY ($NiPS_3$ [32–35]), and Ising ($FePS_3$ [28,31,35–38]) types, which depend solely on the exchange coupling energies between transition metal ions in the lattice. The magnetic structures and magnetic ordering temperatures are mediated by the strength and correlation between the first-, second-, and third-neighbor interactions of the in-plane metal atoms [28,29]. The interlayer exchange coupling, though relatively weak, plays a crucial role in the formation of the magnetic order along the *c* axis [34].

Although the $MPX_3$ family has diverse magnetic properties, none of the explored $MPX_3$ materials exhibit spin frustration or incommensurate spin arrangements. In this report, for the first time, we explore a weak spin frustration system in this family, $CrPSe_3$. It crystallizes in the monoclinic structure of C2/*m* symmetry and orders antiferromagnetically at the Néel temperature $T_N \sim 126$ K. The large frustration factor, $f = |\theta_{CW}|/T_N$ (~3) implies that $CrPSe_3$ has weak spin-frustration. The modulated spin density wave and incommensurate AFM order in $CrPSe_3$, revealed by neutron diffraction experiments, bring novel prospects for further search of related compounds for studies of magnetic frustration effects and their interplay with physical properties of the layered 2D vdW compounds.

## II. EXPERIMENTAL PROCEDURE

As the first stage of synthesis procedure, a polycrystalline powder of $CrPSe_3$ was synthesized by a solid-state reaction. Reagents of Cr (powder, 200 mesh, Alfa Aesar, 99.99%), P (red, lumps, Alfa Aesar, 99.99%), and Se (crushed granules, Alfa Aesar, 99.999%) were ground together under ambient conditions. Approximately 2 g of materials with their stoichiometric ratio Cr:P:Se = 1:1:3 were pressed as a pellet at 0.1 MPa pressure, followed by sealing in a quartz ampule (22×25×100 mm) under high vacuum ($<5\times10^{-3}$ Torr). The sealed ampule was kept in a box furnace at 700 $^{o}$C with a ramp rate of 1.6 $^{o}$C per min for a 10-day holding time, followed by furnace cooling. At the second stage, the single crystals were prepared using this powder. This pre-reacted material (1 g) was sealed in a quartz ampoule (22×25×230 mm) along with $I_2$ (~50 mg) and was kept in a two-zone furnace at 800–700 °C with a 10-day holding time. Small crystals of a few millimeters were obtained in the cool zone.

Powder XRD measurements were conducted using a Rigaku Smart Lab diffractometer with Cu Kα radiation (λ = 1.5418 Å). Le-Bail refinement analysis was carried out using Jana2006 [39]. XPS data were collected using a VG Microtech ESCA2000 X-ray photoelectron spectrometer using a monochromatic Al-$K_\alpha$ X-ray source with 1486.7 eV energy.

The magnetic susceptibility as function of temperature and magnetic field was measured using the vibrational sample magnetometer of a Quantum Design Physical Properties Measurement System. The sample loaded at room temperature was cooled to a low temperature (2 K) without applying magnetic field (zero field cooling, ZFC). At 2 K, a field of 0.1 T was applied, and the susceptibility was measured on heating up to 300 K. The same magnetic field was applied during the second cooling process (field cooling, FC) and data were collected while heating to 300 K. The magnetic moment versus field measurements were conducted at 2 K in the fields range from -14 to 14 T and vice versa.

The room temperature and temperature dependent Raman spectra were collected using Horiba iHR 550 spectrometer with 514 nm wavelength laser and optical setup of own construction with the 532 nm wavelength laser in the inverted mode [40], respectively. The Montana system (Cryostation s50) was used to provide low temperatures in the range between 20 and 250 K. The sample was illuminated by low laser power (~500 μW) to protect it from unintentional laser effects and the Raman spectra were collected in reflected geometry. To acquire the Raman spectra, the collected signals were guided to a 50-cm-long spectrometer equipped with a cooled charge-coupled device (CCD) using an optical fiber with a core diameter of 150 μm.

The neutron powder-diffraction (NPD) measurements in the temperature range of 25–300 K were performed using the DN-12 diffractometer (IBR-2 pulsed reactor, JINR, Russia) [41]. The NPD patterns were collected at scattering angle of $2\theta = 90°$ with a resolution of $\Delta d/d = 0.015$. The typical data-collection time for each temperature was 12 h. The experimental data were analyzed by the Rietveld method using the FullProf program [42].

## III. RESULTS AND DISCUSSION

### A. Crystal structure, chemical composition, and macroscopic magnetic properties

Figure 1(a) depicts the room-temperature XRD patterns of $CrPSe_3$. The data refinement indicates that the crystal structure of the synthesized $CrPSe_3$ material belongs to the C2/*m* space group. No signatures for additional impurity phases in our synthesized $CrPSe_3$ sample were found. It is worth noting that the structure of monolayer $CrPSe_3$ is distinctive from other isostructural members of the $MPX_3$ family with the C2/*m* space group. Figure 1(b) shows the top (along the z-direction or c*) and side (along the y-direction or b*) views. In the monolayer structures of the $MPX_3$ members (M = Cr, Mn, Fe, Co, and Ni; X = S, Se), metal atoms are located in a honeycomb lattice, and P-P bonds are almost perpendicular to the metal plane. In contrast, the P-P bond of $CrPSe_3$ was tilted (~ 14.5°) in the direction perpendicular to the Cr plane (Table II). Furthermore, the significant difference in the M-M, P-$Se_1$, and P-$Se_2$ bond lengths (denoted in Fig. 1(c) and Table II) creates the $C_{2h}$ point group in monolayer $CrPSe_3$ instead of the $D_{3h}$ of other $MPX_3$ [34,43].

The chemical bonding characteristics of $CrPSe_3$ were investigated by a high-resolution XPS spectroscopy measurements performed for each atom (Fig. 1(d)) - Cr *3p* (top), P *2p* (middle), and Se *3d* (bottom). Deconvolution analysis of the bonding modes revealed the main peaks at 574.41 eV (Cr $2p_{3/2}$) and 585.13 eV (Cr $2p_{1/2}$). In addition, small shadow peaks developed at 575.79 eV and 583.85 eV imply a presence of weak nonstoichiometry in the studied material. Similarly, shadow peaks were also observed in the XPS spectrum of P-*2p* at 132.45 eV and Se-*3d* at 55.85 eV in addition to the main peaks (e.g., P $2p_{3/2}$ (130.40 eV) and P $2p_{1/2}$ (131.37 eV); Se $3d_{5/2}$ (53.32 eV) and Se $3d_{3/2}$ (54.82 eV)). Quantitative analysis results in the composition of $Cr_{0.9}P_{1.1}Se_3$, instead of ideal $Cr_1P_1Se_3$.

We further investigated the magnetic properties of $CrPSe_3$. Magnetic susceptibility measurements ($\chi$) revealed a magnetization drop below 150 K, indicating the formation of an antiferromagnetic order (Fig. 2(a)). The Néel temperature $T_N \approx 126$ K was extracted from the analysis of the $d\chi/dT$ derivative. The Curie-Weiss temperature ($\theta_{CW}$) and effective magnetic moment ($\mu_{eff}$) were determined from the inverse susceptibility $1/\chi - T$ plot in the 200–250 K range (Fig. 2(b)). The obtained value of $\theta_{CW} \sim$ -371 K is the largest one among the members of the $MPX_3$ family (Table I). It implies a presence of strong short-range antiferromagnetic interactions in $CrPSe_3$ in the paramagnetic region. The effective magnetic moment of $CrPSe_3$, $\mu_{eff} \sim 4.6\ \mu_B$, corresponds to the high-spin state (S = 2) of $Cr^{2+}$ ions**.** In addition, magnetic field dependent magnetization isotherms (Fig. 2(c)) were measured at 2 K in fields range up to 14 T. The non-linear behavior beyond $\pm$10 T could be attributed to spin-flop effects at high fields, similar to that found in $\alpha$-$RuCl_3$ [44].

The Raman measurements were conducted on exfoliated flakes of $CrPSe_3$ on a $SiO_2$/Si substrate obtained by the scotch-tape method. At room-temperature, eight vibrational modes were detected in the 100-350 $cm^{-1}$ range (Fig. 3(a)). The origin of observed modes is likely associated with the normal modes of the $[P_2Se_6]^{4+}$ dimers, which are similar to those of $FePS_3$ and $FePSe_3$ [36,43,45,46]. Six Raman vibrational modes were observed in $FePS_3$ at room temperature in the higher frequency region above 100 $cm^{-1}$. Nevertheless, more Raman peaks appear in $CrPSe_3$ since the point group symmetry order is lower in $CrPSe_3$ compared to that in $FePS_3$. The number of peaks is unchanged in the temperature-dependent Raman spectra from 20 K to 225 K (Fig. 3(b)), denoting the stability of the monoclinic phase in $CrPSe_3$ and the absence of the spin-order induced vibrational modes. This is contrasted to the appearance of extra Raman peaks below $T_N$ in $FePS_3$ [36,45,47]. Nevertheless, anomalous temperature-dependent shift of the vibrational peak located at ~208 $cm^{-1}$ (assigned as $P_6$) was clearly revealed (Fig. 3(c)). This peak is associated with the $A_{1g}$ mode of the $P_2Se_6$ group (inset of Fig. 3(c)) of $MPX_3$ family, involving partially the Cr-Cr atoms vibrations due to the lower point group symmetry of $CrPSe_3$. A change of the curve slope in the vicinity of $T_N$ = 126 K implies a presence of pronounced spin-phonon coupling, which was also previously found in vdW magnets like $CrBr_3$ and $Cr_2Ge_2Te_6$ [20,48]. We did not analyze other detected Raman lines in detail because of their lower intensities.

### B. Long range magnetic order and generation of spin density wave

To investigate temperature evolution of the crystal and magnetic structure of $CrPSe_3$, the neutron powder diffraction (NPD) patterns were measured on cooling to 25 K, as shown in Fig. 4a. The Rietveld refinements of the experimental data revealed that monoclinic symmetry of the crystal structure remains unchanged at low temperatures, in consistence with Raman spectroscopy results (Fig. 4(b)). Below 120 K, appearance of two magnetic peaks at $d_{hkl}$ = 6.62 and 7.63 Å was observed. From the Rietveld analysis of the NPD pattern at 25 K, we found that these peaks are associated with formation of the long-range incommensurate spin density wave (SDW) antiferromagnetic order with the propagation vector $\boldsymbol{k}$ = (0, 0.40, 0), which can be considered as the modulated zigzag type (Fig. 4(b)). Furthermore, the orientation of the ordered magnetic moments of Cr in the SDW chain is perpendicular to *ab* plane (Fig. 4(c)). Their values at $T$ = 25 K, $\mu_{\text{eff}}$ = 3.2 $\mu_B$, are close to those expected for the high-spin state of $Cr^{2+}$ ions. A reduction in the local ordered magnetic moment compared to effective paramagnetic moment is typical for 2D vdW magnetic materials [18,49–51] and may be related to local magnetic disorder caused by low dimensionality effects.

Compared to other previously known magnetic orders of $MPX_3$ (M = Fe, Mn, Ni, X = S, Se) systems, which are listed in Table II, $CrPSe_3$ becomes the unique one with the SDW spin arrangement. To understand the nature of the SDW state in $CrPSe_3$, the parameter $f = |\theta_{CW}|/T_N$, which represents the degree of magnetic frustration, was considered. Typically, values $f > 5$ correspond to strongly frustrated systems with a suppressed long-range magnetic order [52]. The lower values $f \sim 3\text{-}4$ correspond to weakly frustrated systems, where several competitive magnetic orders could be developed, like λ-$MnO_2$ ($f = 3.2$) [53] and $TbMnO_3$ ($f = 4.2$) [54]. The $CrPSe_3$ has $f$ value of 2.9, the largest one in the $MPX_3$ family (Table I). This value is in the range of other weakly frustrated systems such as λ-$MnO_2$ and $TbMnO_3$ ($TbMnO_3$ has several transitions, and we consider the lowest transition temperature (~10 K) instead of using those of the first incommensurate transition (~37 K). One of the recently proposed scenarios of the SDW state formation in these weakly frustrated systems points to the superposition of two magnetic sets of opposite chirality [51], and it may be also applicable for the present case of $CrPSe_3$.

In conclusion, we successfully synthesized $CrPSe_3$ material by chemical vapor transport method. We found that $CrPSe_3$ is a weakly frustrated antiferromagnet with the frustration parameter $f \sim 3$, giving rise to formation of the long-range incommensurate spin-density wave state. Our finding is the first report for a weakly frustrated two-dimensional van der Waals insulator, enriching the physics of magnetism at the 2D limit.

## ACKNOWLEDGEMENTS

This study was supported by the Institute for Basic Science (Grant number IBS-R011-D1) and Advanced Facility Center for Quantum Technology. Tuan Anh Tran acknowledged for the Vietnam Ministry of Education and Training grant No. B2022-SPK-07.

## AUTHOR CONTRIBUTIONS

M. B., N. T. D., and T. A. T. contributed equally to this work.

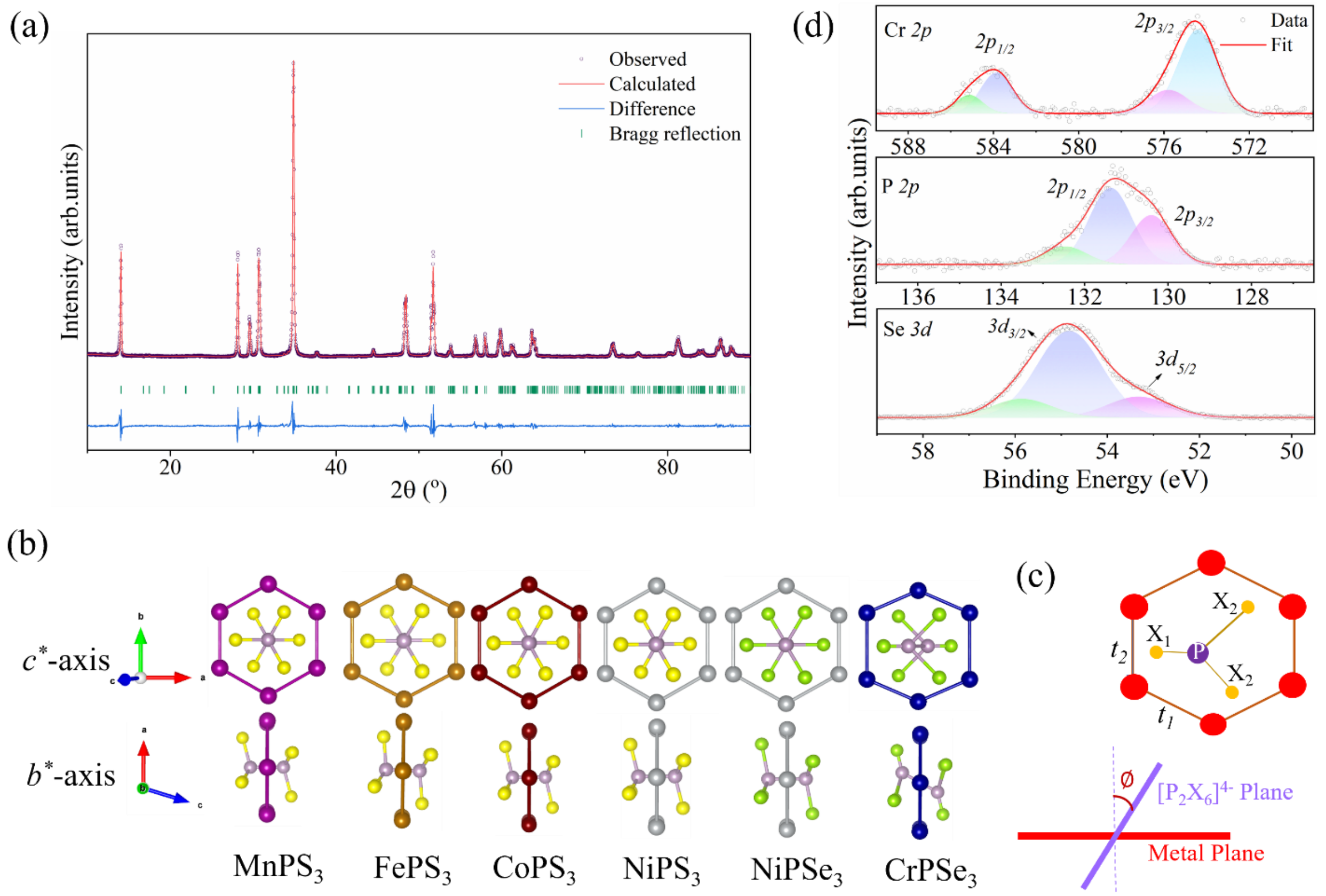

FIG. 1. (a) Room temperature x-ray diffraction pattern of the synthesized $CrPSe_3$ material. The experimental points, calculated profile, Bragg reflections positions and difference curve are shown. (b) Monolayer structure comparison among members of $MPX_3$ (M = Cr, Mn, Fe, Co, and Ni; X = S, Se) family. The metal atoms are located in a honeycomb lattice and ($[P_2Se_6]^{4-}$) dimer is located in a perpendicular plane to metal plane. (c) The layout of the M-M distances ($t_1$, $t_2$) in honeycomb lattice, the P-P, and P-X distances, which values are listed in Table II. (d) High-resolution x-ray photoelectron spectra and peak deconvolution of Cr *2p*, P *2p*, and Se *3d*.

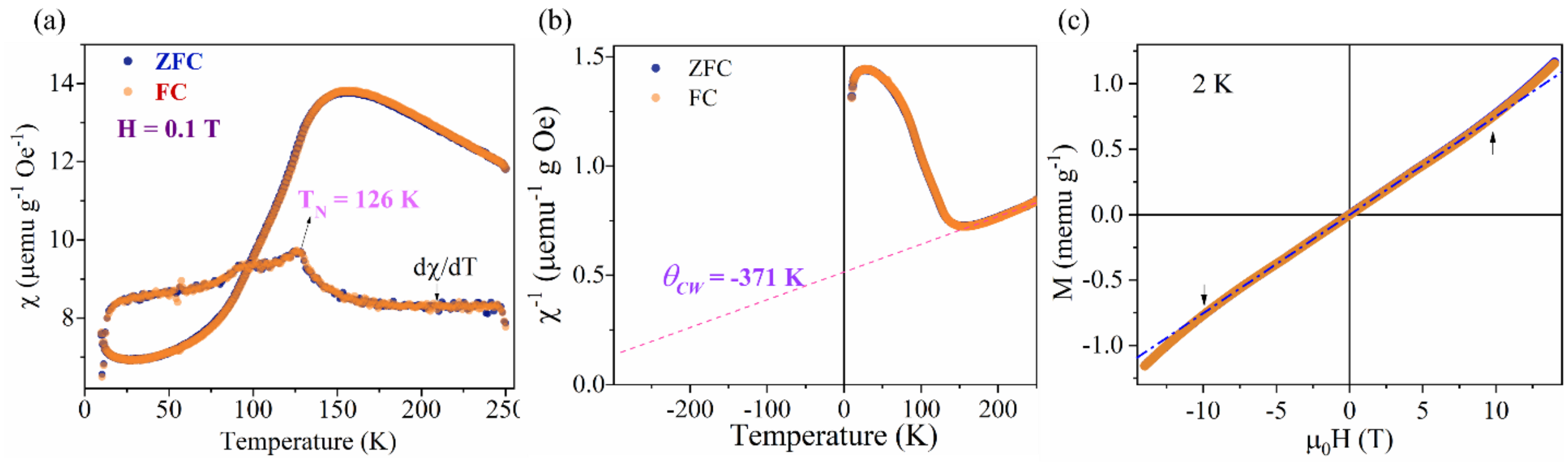

FIG. 2. (a) Magnetic susceptibility of $CrPSe_3$ powder as a function of temperature measured at H = 1 kOe, the $\chi$ - T curves, blue color for zero field cooling (ZFC) and orange color for field cooling (FC), showing antiferromagnetic transition with the Néel temperature $T_N \sim 126$ K (extracted from the dχ/dT versus T curve). (b) $\chi^{-1}$ versus T curve, high Curie-Wiess temperature indicates strong exchange interactions in paramagnetic region (above $T_N$). (c) Field dependent magnetization measured at 2 K.

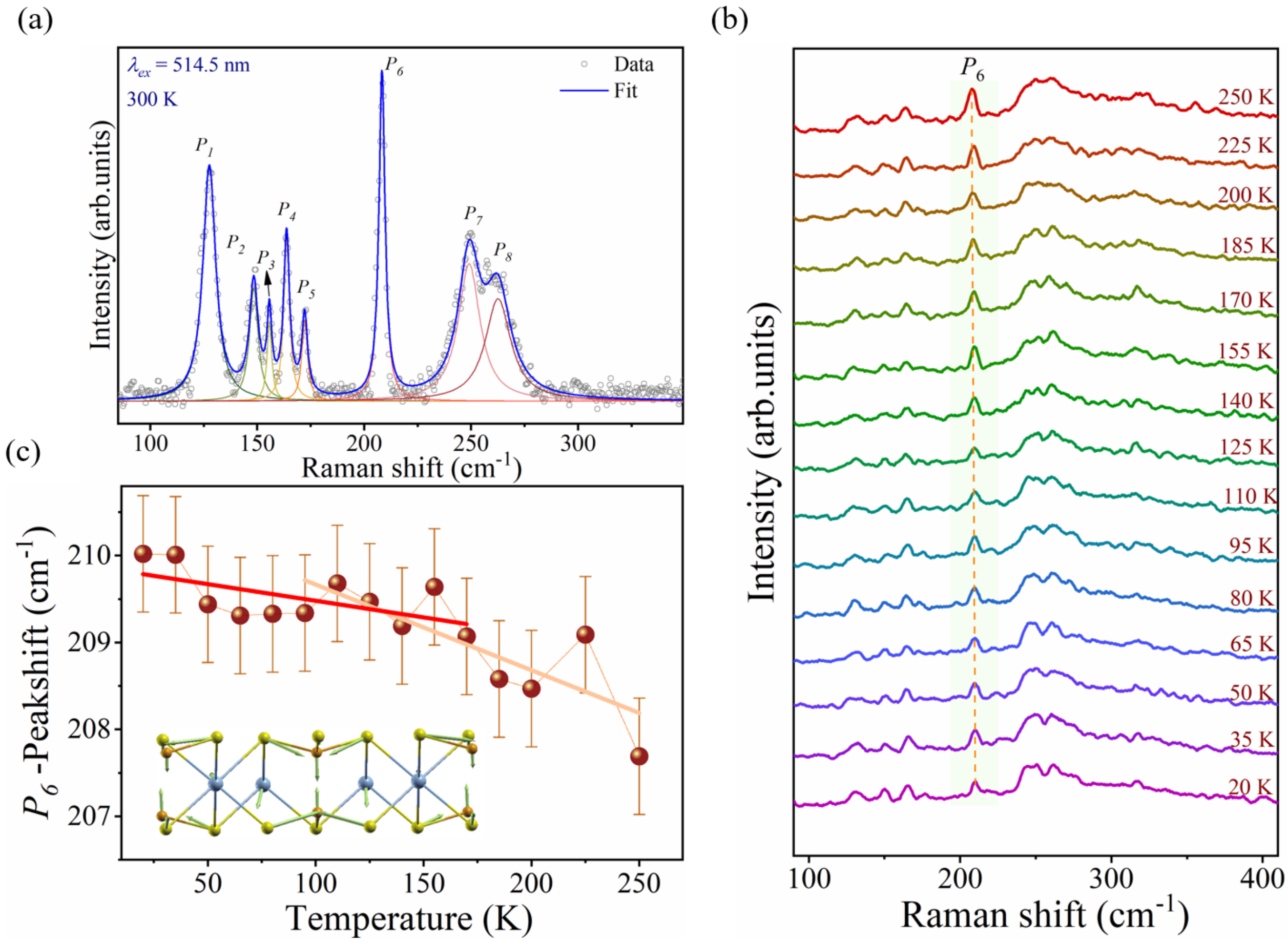

FIG. 3. (a) Raman spectra of $CrPSe_3$ showing prominent vibrational modes at room temperature, (b) Temperature-dependent Raman spectra at 20–250 K indicate unchanged crystal structure throughout the temperature region as indicated by the consistent vibrational modes. (c) The evolution of the $P_6$ peak with temperature, revealing two different slopes at a critical temperature of 125 K.

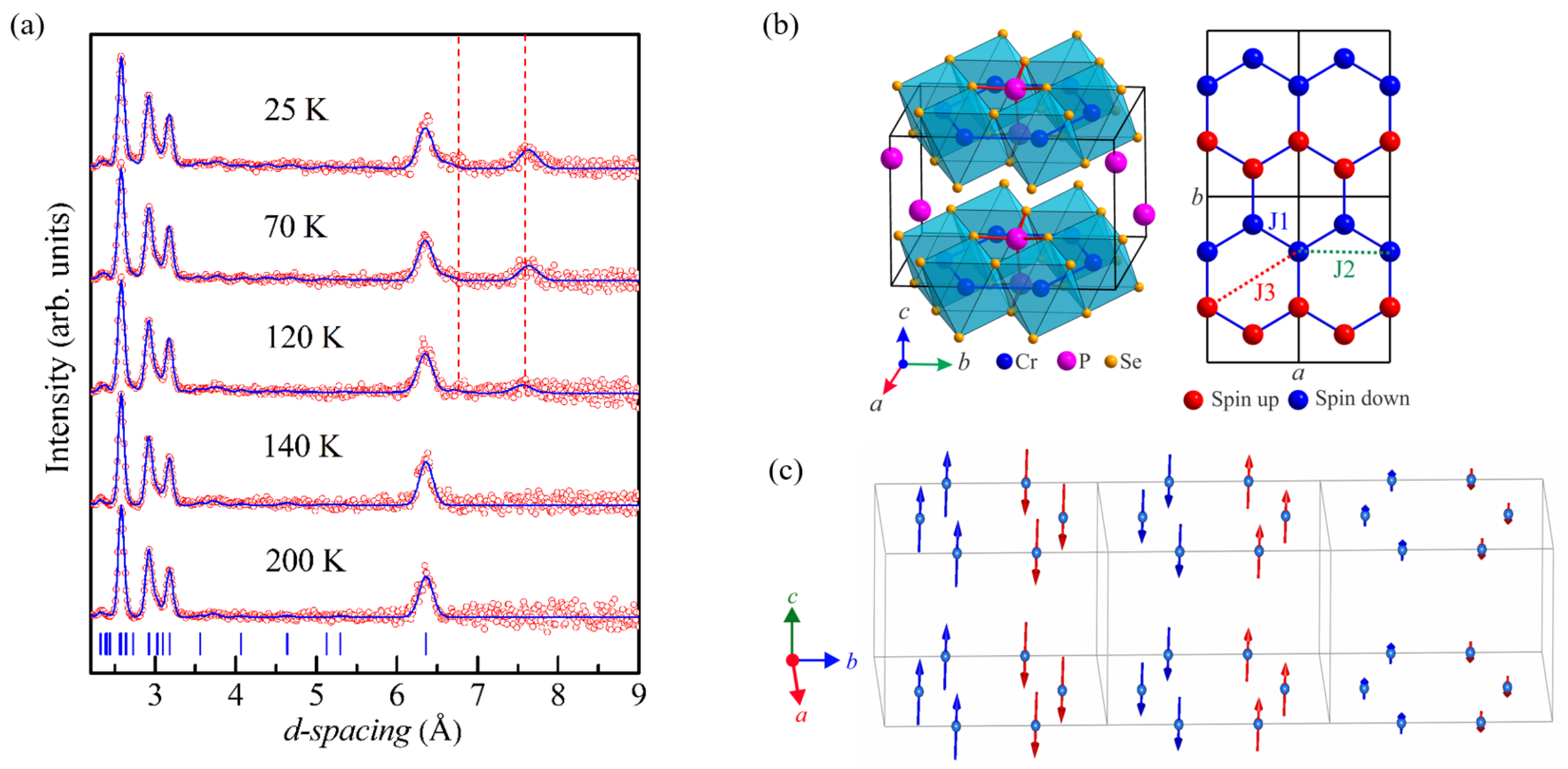

FIG. 4. (a) The powder neutron diffraction patterns measured from 25 K to 200 K. The experimental points, calculated profiles and calculated nuclear peaks positions are shown. The appearance of additional magnetic peaks at 6.7 Å and 7.5 Å below 140 K indicates the formation of antiferromagnetic order consistent with magnetic susceptibility. (b) The monoclinic crystal structure extracted from Rietveld refined data (*left*) and spin configuration (*right*), $J_1$, $J_2$, and $J_3$ are exchange interaction energies of first, second, third nearest neighbors, respectively. (c) The incommensurate spin density wave magnetic structure formed in $CrPSe_3$. The values of the ordered magnetic moments vary along the b-axis.

TABLE I. Crystal structure, antiferromagnetic transition temperatures, effective paramagnetic and ordered magnetic moments, types of magnetic order, and in-plane exchange energies of $MPX_3$ family of materials in comparison with $CrPSe_3$.

| **Compound** | **Crystal Structure (Space group)** | **Spin order** | $T_N$ (K) | $\boldsymbol{\theta}_{CW}$ (K) | **f** = $\frac{\lvert\theta_{CW}\rvert}{T_N}$ | $\boldsymbol{\mu_{eff}\ (\mu_B)}$ | |
|---|---|---|---|---|---|---|---|
| | | | | | | ***χ(T)-T*** | ***NPD*** |
| $NiPSe_3$ [29,55] | Monoclinic (C2/*m*) | zigzag | 206 | - | - | - | - |
| $NiPS_3$ [28,33,35] | Monoclinic (C2/*m*) | zigzag | 155 | -254 | ~1.6 | 3.9 [56] | 1.06 [57] |
| $CoPS_3$ [50,58] | Monoclinic (C2/*m*) | zigzag | 122,132 | -116, -223 | 0.9,1.9 | 4.9 [50] | 4.9 [56] |
| $FePSe_3$ [12,18] | Rhombohedral (R-3) | zigzag | 112 | -9 | ~0.1 | 5.2 [18] | 4.2 [18] |
| $FePS_3$ [13,28] | Monoclinic (C2/*m*) | zigzag | 118,119 | -112 | ~1.0 | 4.9 [56] | 4.5 [49] |
| $MnPSe_3$ [12,18] | Rhombohedral (R-3) | Néel | 74 | -146 | ~2.0 | 5.9 [18] | 3.6 [18] |
| $MnPS_3$ [13,19,28] | Monoclinic (C2/*m*) | Néel | 78,80 | -160 | ~2.0 | 5.98 [56] | 4.1 [59] |
| **$\mathbf{CrPSe_3}$ [This work]** | **Monoclinic (C2/*m*)** | **zigzag** | **126** | **-371** | **2.9** | **4.6** | **3.2** |

TABLE II. The structural parameters characterizing monolayer geometry among members of the $MPX_3$ family compared to $CrPSe_3$, including lattice parameters, M-M distance on hexagonal lattice ($t_1$ and $t_2$ shown in Fig. 1(c)), P-P, P-X distance, $[P_2X_6]^{4-}$ dimer deviation angle from a perpendicular to metal plane (∅).

| Compound | Lattice parameters | | M-M distance | Bond lengths (P-P & P-X) | Deviation (∅) |
|---|---|---|---|---|---|
| $MnPS_3$ [60] | $a$ = 6.0770 Å<br>$b$ = 10.5240 Å<br>$c$ = 6.7960 Å | $\alpha=\gamma$ = 90°,<br>$\beta$ = 107.3500°<br>V = 414.8583 Å$^3$ | $t_1$ = 3.5005 Å<br>$t_2$ = 3.5239 Å | $d_{P-P}$ = 2.1874 Å<br>$d_{P-S1}$ = 2.0342 Å<br>$d_{P-S2}$ = 2.0296 Å | 0.119° |
| $FePS_3$ [49] | $a$ = 5.9400 Å<br>$b$ = 10.2600 Å<br>$c$ = 13.2000 Å | $\alpha=\gamma$ = 90°,<br>$\beta$ = 108.3000°<br>V = 763.7806 Å$^3$ | $t_1$ = 3.3641 Å<br>$t_2$ = 3.5500 Å | $d_{P-P}$ = 2.0566 Å<br>$d_{P-S1}$ = 2.0318 Å<br>$d_{P-S2}$ = 2.0744 Å | 2.058° |
| $CoPS_3$ [37] | $a$ = 5.9010 Å<br>$b$ = 10.2220 Å<br>$c$ = 6.6580 Å | $\alpha=\gamma$ = 90°,<br>$\beta$ = 107.1700°<br>V = 383.7122 Å$^3$ | $t_1$ = 3.3874 Å<br>$t_2$ = 3.4469 Å | $d_{P-P}$ = 2.1501 Å<br>$d_{P-S1}$ = 2.0392 Å<br>$d_{P-S2}$ = 2.0556 Å | 0.223° |
| $NiPS_3$ [61] | $a$ = 5.8120 Å<br>$b$ = 10.0700 Å<br>$c$ = 6.6320 Å | $\alpha=\gamma$ = 90°,<br>$\beta$ = 106.9800°<br>V = 371.2293 Å$^3$ | $t_1$ = 3.3525 Å<br>$t_2$ = 3.3634 Å | $d_{P-P}$ = 2.1693 Å<br>$d_{P-S1}$ = 1.9851 Å<br>$d_{P-S2}$ = 1.9890 Å | 0.005° |
| $NiPSe_3$ [55] | $a$ = 6.1604 Å<br>$b$ = 10.6768 Å<br>$c$ = 6.9323 Å | $\alpha=\gamma$ = 90°,<br>$\beta$ = 107.1791°<br>V = 435.6165 Å$^3$ | $t_1$ = 3.5574 Å<br>$t_2$ = 3.5586 Å | $d_{P-P}$ = 2.2413 Å<br>$d_{P-Se1}$ = 2.2466 Å<br>$d_{P-Se2}$ = 2.2468 Å | 0.055° |
| **$CrPSe_3$ [This work]** | **$a$ = 6.1246 Å**<br>**$b$ = 10.6705 Å**<br>**$c$ = 6.6950 Å** | **$\alpha=\gamma$ = 90°,**<br>**$\beta$ = 107.8560°**<br>**V = 416.4593 Å$^3$** | **$t_1$ = 3.3428 Å**<br>**$t_2$ = 3.9948 Å** | **$d_{P-P}$ = 2.2255 Å**<br>**$d_{P-Se1}$ = 2.0307 Å**<br>**$d_{P-Se2}$ = 2.2995 Å** | **14.472°** |